# Effects of bacterial density on growth rate and characteristics of microbial-induced CaCO$_3$ precipitates: a particle-scale experimental study

Yuze Wang[1]*; Kenichi Soga[2]; Jason T. DeJong[3]; Alexandre J. Kabla[4]

[1]Department of Ocean Science and Engineering, Southern University of Science and Technology, 518055, People's Republic of China
[2]Department of Civil and Environmental Engineering, University of California, Berkeley, CA 94720, United States
[3]Department of Civil and Environmental Engineering, University of California, Davis, CA 95616, United States
[4]Department of Engineering, University of Cambridge, Cambridge, CB2 1PZ, United Kingdom
*Correspondence: wangyz@sustech.edu.cn; ORCID: 0000-0003-3085-5299

**Abstract** Microbial-Induced Carbonate Precipitation (MICP) has been explored for more than a decade as a promising soil improvement technique. However, it is still challenging to predict and control the growth rate and characteristics of CaCO$_3$ precipitates, which directly affect the engineering performance of MICP-treated soils. In this study, we employ a microfluidics-based pore scale model to observe the effect of bacterial density on the growth rate and characteristics of CaCO$_3$ precipitates during MICP processes occurring at the sand particle scale. Results show that the precipitation rate of CaCO$_3$ increases with bacterial density in the range between $0.6\times10^8$ and $5.2\times10^8$ cells/ml. Bacterial density also affects both the size and number of CaCO$_3$ crystals. A low bacterial density of $0.6\times10^8$ cells/ml produced $1.1\times10^6$ crystals/ml with an average crystal volume of 8,000 µm$^3$, whereas a high bacterial density of $5.2\times10^8$ cells/ml resulted in more crystals ($2.0\times10^7$ crystals/ml) but with a smaller average crystal volume of 450 µm$^3$. The produced CaCO$_3$ crystals were stable when the bacterial density was $0.6\times10^8$ cells/ml. When the bacterial density was 4-10 times higher, the crystals were first unstable and then transformed into more stable CaCO$_3$ crystals. This suggests that bacterial density should be an important consideration in the design of MICP protocols.

## INTRODUCTION

Microbial-Induced Calcium Carbonate Precipitation (MICP) has been extensively investigated for applications such as ground improvement, soil liquefaction mitigation, dam safety control, prevention of soil erosion, and slope stabilisation (van Paassen, 2009; DeJong *et al.*, 2013; Martinez *et al.*, 2013; Montoya *et al.*, 2013; Jiang *et al.*, 2017). The CaCO$_3$ precipitates fill soil pores and bond soil particles, which consequently increase the strength and stiffness, and reduce the permeability of the soil matrix (Stocks-Fischer *et al.*, 1999; DeJong *et al.*, 2006). Several types of bacterial activities including ureolysis, denitrification and sulfate reduction can result in MICP (DeJong *et al.*, 2010), and among those the ureolysis-based process has been most widely studied. Ureolysis-driven MICP involves urea hydrolysis by the urease enzyme produced by active microorganisms (Equation 1), resulting in the generation of calcium carbonate (CaCO$_3$) in the soil matrix (Equation 2).

$$CO(NH_2)_2 + 2H_2O \xrightarrow{Urease} 2NH_4^+ + CO_3^{2-}$$  Eq. 1

$$Ca^{2+} + CO_3^{2-} \rightarrow CaCO_3(s)$$  Eq. 2

Ureolysis and CaCO$_3$ precipitation are the two key processes involved in the ureolysis-driven MICP process. Understanding the kinetics of these two processes is essential for designing MICP protocols. The kinetics of ureolysis is normally assessed by the increase in solution conductivity due to the hydrolysis of urea (Whiffin *et al.*, 2007; Lauchnor *et al.*, 2015). The kinetics of CaCO$_3$ precipitation can be assessed by the decrease in Ca$^{2+}$ concentration (Stocks-Fischer *et al.*, 1999). Recently, the kinetics of CaCO$_3$ at the crystal size level have also been studied by using optical microscopes to observe CaCO$_3$ crystals produced in liquid medium in petri dishes (Zhang *et al.*, 2018), on glass slides (Wang *et al.* 2019b) or in microfluidic chips (Wang *et al.* 2019a,b; Kim *et al.*, 2020), as well as crystals produced on solid agar pads (Zhang *et al.*, 2018). Crystals grew steadily to diameters of 20 μm and 50 μm within 40 minutes when MICP occurred in a liquid medium placed in a petri dish or on an agar pad containing Ca$^{2+}$ urea, respectively (Zhang *et al.*, 2018). In comparison, the use of a microfluidic-based porous model is considered to be a more appropriate approach to study the kinetics of MICP as it more closely mimics real MICP conditions occurring in the pore fluid of a porous soil matrix where bacterial cells move freely and the cementation solution can be injected multiple times (Wang *et al.*, 2019a,b).

In addition to the CaCO$_3$ precipitation kinetics, the properties of CaCO$_3$ precipitates also need to be considered in a MICP protocol design. Larger crystals that bond soil particles more sufficiently may increase soil strength more effectively (Cheng *et al.*, 2017). By conducting soil column experiments and by using a scanning electronic microscopy to scan the samples after MICP treatment, it was found that the concentration of cementation solution and ureolysis activity affected the size and number of CaCO$_3$ crystals after MICP treatment (Al Qabany & Soga, 2013; Cheng *et al.*, 2017). Al Qabany & Soga (2013) found that when the total treatment duration and the total amount of cementation solution were constant, the use of higher concentrations of cementation solution produced larger CaCO$_3$ crystals. Cheng *et al.* (2017) showed that higher bacterial activities tend to produce smaller CaCO$_3$ crystals at the end of MICP treatment.

Because CaCO$_3$ crystals were only observed after MICP treatment by using a scanning electronic microscopy (Al Qabany & Soga, 2013; Cheng *et al.*, 2017), the kinetics and characteristics of microbial-induced CaCO$_3$ precipitation were not fully understood. Wang *et al.* (2019a) designed and fabricated a microfluidic chip etched with porous models and used it to observe the MICP process under conditions that resemble soil assemblies. The advantage of this method is that the density of bacteria and the main parameters of CaCO$_3$ crystals such as size, shape and number can be quantified during the whole MICP process (Wang *et al.*, 2019 a,b). Wang *et al.* (2019b) found that when the bacterial activity and concentration of cementation solution were the same, longer injection intervals (23-25 hours compared to 3-5 hours) produced larger and fewer CaCO$_3$ crystals. This was because when the interval was longer, the smaller and less stable crystals dissolved while the larger and more stable crystals continued to grow (Wang *et al.*, 2019b).

Therefore, it is essential to investigate the processes of MICP rather than only the MICP properties after MICP treatment to have a better understanding of the kinetics and properties of MICP. Due to the fact that bacterial density has a direct effect on the ureolysis activity (Lauchnor *et al.*, 2015) and a large range of bacterial densities have been used in MICP studies (Al Qabany *et al.*, 2012; Cheng *et al.*, 2017), it is essential to know the quantity and activity of the bacteria injected into the soil to design robust MICP treatment protocols. In this study, microfluidic experiments were conducted to observe both the growth kinetics and characteristics of microbial-induced CaCO$_3$ crystals under conditions where the bacterial densities varied. The test results were used to investigate the effects of bacterial density on the kinetics and characteristics of CaCO$_3$ precipitation and to provide suggestions for MICP treatment protocols.

**MATERIALS AND METHODS**

**Microfluidic chip experimental setup and data acquisition**

The microfluidic chip used in this study was designed based on a cross-sectional image of a solidified and sectioned 3D Ottawa 30–50 sandy soil specimen (image provided courtesy of Dr. David Frost) and the fabrication of the microfluidic chip was conducted following the standard photolithography techniques by using polydimethylsiloxane (PDMS) and glass. The design and fabrication methods of the microfluidic chips were discussed in Wang *et al.* (2019a). The experimental setup is shown in Figure 1.

During the experiments, all images were acquired with an Axio Observer Z1 research microscope. The microscope is equipped with an automated stage (Prior Scientific Instrument), a grayscale camera (Hamamatsu C11440-22CU), and a light source connected to a computer and controlled by Zeiss AxioVision image analysis software. Images were captured using phase field illumination and 10× inverted objectives (with image resolutions of 0.65 µm/pixel). In the images, bacterial cells appeared as black dots, the $CaCO_3$ precipitates appeared as white dots, and the microfluidic chip appeared as light to dark grey background.

**Preparation of bacterial suspension and cementation solution**

*Sporosarcina pasteurii* (DSM 33), a ureolytic bacterial strain, was used in the experiments. Bacterial cells from the glycerol stock (Wang *et al.*, 2019b) were grown in ATCC 1376 $NH_4$-YE agar medium (20 g/L yeast extract, 10 g/L ammonium sulphate, 20 g/L agar, and 0.13 M Tris base) for 48 hours at 30°C. Subsequently, several colonies on the agar plate were transferred to a $NH_4$-YE liquid medium containing the same components without agar and cultivated in a shaking incubator for 24 hours at 30°C and at a shaking rate of 200 rotations per minute (rpm) to obtain a bacterial suspension with an optical density measured at a wavelength of 600 nm ($OD_{600}$) of around 3.0. A more detailed description of the preparation of the bacterial suspension can be found in Wang *et al.* (2019b). The cementation solution contained 0.25 M of $CaCl_2$, 0.375 M of urea and 3 g/L of nutrient broth.

**Staged injection procedure of cementation solution to produce $CaCO_3$ precipitates**

A staged injection procedure was applied to the MICP treatment process. After bacterial injection into a microfluidic chip and the subsequent 2 hour bacterial settling period, 1.25 pore volumes (PV) of cementation solution was injected into the microfluidic chip at an injection flow rate of 5.6 PV/hr. Subsequent injections of cementation solution were conducted at 24-hour intervals after the previous injection. In total, 12 injections of cementations solution were applied. The injection volume of cementation solution, the flow rate, and the number of injections of cementation solution were chosen based on findings presented in Wang *et al.* (2019a, b). Time-series images were captured at 15 minutes intervals after the completion of each of the cementation solution injections. The main experimental parameters in these protocols are summarised in Table 1.

**Quantification of bacterial density and activity**

Bacterial suspensions with five different densities were prepared to correlate with bacterial optical density. The $OD_{600}$ was measured to quantify the bacterial density before the injection of 1.5 PV into each of five microfluidic chips at an injection flow rate of 56 PV/h. The bacterial suspensions were prepared by diluting the bacterial suspension with $OD_{600}$ of 3.0 using autoclaved $NH_4$-YE liquid medium at volume proportions V_bacterial suspension : V_$NH_4$-YE liquid medium of 1:14, 1:5, 1:2, 2:1 and 1:0. Injecting 1.5 PV of bacterial suspension at a flow rate of 56 PV/h resulted in a homogeneous distribution of bacteria after injection (Wang *et al.*, 2019a). The bacterial density was quantified based on the images taken at the centre of the microfluidic chips. Bacteria were given 10 mins to settle to the bottom zone of the microfluidic chip before obtaining accurate microscope images since the microscope focal length depth range was 17 μm, whereas the depth of the microfluidic chip was 50 μm. The bacterial cells doubled in number in about 2 hours due to *in situ* growth during bacterial settling (Wang *et al.*, 2019a). The effect of bacterial growth on the bacterial density changed during the first 10 minutes, but this was neglected due the short period. Because bacterial size affects the reading of $OD_{600}$ value (Zapata & Ramirez-Arcos, 2015), the sizes of the bacteria were also measured to obtain the average bacterial size.

To examine the effect of bacterial density on the rate of ureolysis, a series of batch tests was conducted by varying the bacterial densities in bacteria-urea mixtures. The hydrolysis rate was measured using the conductivity method described by Whiffin *et al.* (2007). The urea concentration in the bacteria-urea mixtures was 1.0 M before the hydrolysis of urea occurred. The bacterial densities in the bacteria-urea mixtures were equivalent to $OD_{600}$ of 0.1, 0.25, 0.46, 0.75, 1.035, 1.38 and 1.73. The conductivity of the mixed content was assessed by using a conductivity meter (FiveGo, Meter Toledo) immediately after the mixing and five minutes after mixing. The ureolysis rate was calculated using Equation 3 (Whiffin *et al.*, 2007). Measurements were performed in triplicates for each of the different media tested, with data presented as mean ± standard error.

$$Ureolysis\ rate(mM/h) = \frac{\Delta Conductivity(\mu S/cm)}{\Delta t(\min)} \times (10^{-3} \times 11.11)(mM/(\mu S/cm)) \times 60(\min/h) \quad \text{Eq. 3}$$

**Quantification of crystal growth rate, size, and quantity**

Several methods were used to quantify $CaCO_3$ crystal characteristics. Mean intensity values of the images were analysed using Zeiss Axio Vision image analysis software and plotted against time to show how the relative areas occupied by the precipitates changed with time. The image intensities could not represent the total crystal volume produced because two-dimensional images cannot capture the entire information of the three-dimensional crystals but could represent the change in crystal properties with time. In addition, crystal diameter has previously been used to quantify the size of $CaCO_3$ crystals (Zhang *et al.*, 2018). During the initial growth stage, most crystals were hemispheres and grew on the surface of the microfluidic channel (Wang *et al.*, 2019; Kim *et al.*, 2020). Therefore, in this study, the volumes of individual crystals were calculated based on their measured diameters. In addition, numbers of crystals were also counted to quantify crystal number.

**RESULTS AND DISCUSSION**

**Bacterial density and bacterial optical density**

The bacterial densities of five bacterial suspensions (cells/ml) were correlated with their initial $OD_{600}$ values to quantify the bacterial density used. Microscope images were taken at the centre of five microfluidic chips containing bacterial suspensions with the initial $OD_{600}$ of 0.2, 0.5, 1.0, 2.0 and 3.0 at 10 minutes after the injection of bacterial suspensions (shown in Figure 2a). Bacterial densities correlated with the initial $OD_{600}$ values of the bacterial suspensions and the results are shown in Figure 2b. A blank sample with no bacterial cells in the bacterial nutrient liquid was used as a baseline against which the $OD_{600}$ was defined as zero. The data shows that bacterial density and $OD_{600}$ are linearly correlated as:

$$Bacterial\ density\ (cells\ per\ ml) = OD_{600} \times 4 \times 10^8 \quad \text{Eq. 4}$$

The $R^2$ values of the all three linear regression lines in Figure 2b are higher than 0.99.

Because bacterial cell size affects the optical density of a bacterial suspension, magnified images of the microfluidic chips containing bacterial suspensions in this study and a previous study (Wang *et al.*, 2019a) are presented to show how the difference in bacterial cell size would affect the optical density of the bacterial suspension (Figure 2c and d). The sizes of bacterial cells in this study are about 3 μm, whereas the ones in Wang *et al.* (2019a) are about 10 μm. Consequently, at the same optical density, the bacterial density in this study is about 3 times higher of that in Wang *et al.* (2019a). Because of the difference in bacterial size, the ureolysis rate may vary when the bacterial optical densities are the same. Therefore, in this study, the bacterial cells were obtained from one batch of bacterial suspension, where the sizes of bacterial cells were relatively consistent, and the bacterial densities were modified to obtain a variety of bacterial activities in terms of ureolysis rate.

**Bacterial density during MICP processes**

Three MICP processes using different bacterial densities were conducted in microfluidic experiments. Optical densities of the bacterial suspensions were measured prior to the injections of bacterial suspensions. Bacterial densities (in cells per ml) of the bacterial suspensions were quantified (i) after injection of bacterial suspension, (ii) after 2 hours of settling, and (iii) after injection of the cementation solution. The results are presented in Table 2. The corresponding microscope images and results are presented in Figures 3a and 3b, respectively. When the initial bacterial densities were $(0.96\pm0.03)\times10^8$, $(3.92\pm0.29)\times10^8$ and $(11.90\pm0.61)\times10^8$ cells/ml, after 2 hours of settling the bacterial densities increased to $(1.92\pm0.13)\times10^8$, $(6.09\pm0.34)\times10^8$ and $(15.4\pm0.88)\times10^8$ cells/ml, respectively, due to bacterial growth *in situ* (Table 2 and Figure 3b). The highest growth rate was obtained when the initial bacterial density was low (about $1\times10^8$ cells/ml); the cell density became about 2 times higher than the initial density after 2 hours. When the initial bacterial density was high (about $4\times10^8$ cells/ml), the bacteria growth rate was about 1.5 times higher than the initial density. When the initial bacterial density was very high (about $12\times10^8$ cells/ml) the growth rate was about 1.25 times higher than the initial densities, which was the lowest among the three cases. The difference in bacterial growth rate might be because the relative abundance of nutrients available to the bacterial population varied depending on the initial bacterial density, with individual bacteria in more concentrated bacterial suspensions being exposed to a smaller share of the total nutrients available.

After the injection of cementation solution, about 30% of the bacteria (approximately $0.6\times10^8$, $2.0\times10^8$ and $5.2\times10^8$ cells/ml for the three cases) remained attached to the inner surface of the microfluidic chip compared to the number of bacteria present after bacterial settling (Table 2 and Figure 3b). The actual bacterial density is expected to be higher than these values since bacterial aggregation also occurred after the cementation solution injection, especially when the bacterial density was high, and the number of bacteria present in bacterial aggregates could not be counted. The percentage of bacteria remaining (30%) is lower than that in Wang *et al.* (2019a), which was 45% after the first injection of cementation solution. This might be because the bacterial settling time was 2 hours in this study, which is lower than the settling time used in Wang *et al.* (2019a), which was 24 hours. In this study, the 2 hours bacterial settling time was used to study the effect of bacterial density on the kinetics and characteristics of MICP.

**Ureolysis rate**

To examine the effect of bacterial density on the rate of ureolysis, a batch test was conducted in which the bacterial densities in bacteria-urea mixtures were varied and the hydrolysis rate was measured by using the conductivity method described by Whiffin *et al.* (2007). The urea concentration in the bacteria-urea mixture was 1.0 M before the hydrolysis of urea occurred (Figure 4). Bacterial density in the range of $0.5\times10^8$ to $4\times10^8$ cells/ml increased the ureolysis rate (Figure 4), while for bacterial densities exceeding around $4\times10^8$ cells/ml the ureolysis rate no longer increased with bacterial density. The linear increase in ureolysis rate associated with bacterial densities in the range of $1\times10^7$ - $2\times10^8$ cells/ml is consistent with the results obtained by Lauchnor *et al.* (2015). The highest previously reported ureolysis rate was measured when bacterial density was $2\times10^8$ cells/ml (Stocks-Fischer et al., 1999; Lauchnor *et al.* 2015). However, the current studyshows that when the bacterial density exceeds $5\times10^8$ cells/ml, the ureolysis rate does not linearly increase with bacterial density, which might be because there could be insufficient nutrients available to sustain bacterial growth at such high bacterial densities .

**Bulk precipitation rate of CaCO$_3$**

The precipitation of CaCO$_3$ with time in the three microfluidic experiments is shown in the images taken at 0.5 hr, 1 hr, 3 hr and 24 hr after the cementation solution injection in Figure 5a. Changes in image intensity with time in the three cases with varied bacterial densities are shown in Figure 5b. The areas in the microfluidic chips occupied by CaCO$_3$ crystals were also different, where a higher bacterial density (e.g. 5.2×10$^8$ cells/ml) resulted in crystals occupying a larger area, as shown by the difference in the image intensities.

We hypothesized that the point when image intensity no longer increases indicated the completion of the CaCO$_3$ precipitation process. The time at which the image intensity stopped increasing in these three cases also varied. The time required for CaCO$_3$ precipitation to complete decreased from 15 hrs to 1.5 hrs when the bacterial densities increased from 0.6 × 10$^8$ to 5.2×10$^8$ cells/ml. The correlation between precipitation rate and bacterial density is shown in Figure 5c. Bacterial density positively affects the overall CaCO$_3$ precipitation rate , and the average precipitation rate in the three cases are 0.016, 0.083 and 0.16 M/h when the corresponding bacterial densities were 0.6×10$^8$, 2.0×10$^8$, and 5.2×10$^8$ cells/ml, respectively.

**Precipitation rates and sizes of individual CaCO$_3$ crystals**

Individual CaCO$_3$ crystals are shown in the magnified images of one of the middle pores inside the microfluidic chips at 0.5, 1, 1.5, 2, 6, 15 and 24 hours after the first cementation injection in Figure 6a. The average crystal volumes calculated based on the measured diameters at an interval of 15 minutes during the first 1.5 hours and over the 24 hours are plotted against time in Figures 6b and 6c, respectively. The average crystal volume data show that, unlike the effect of bacterial density on the overall precipitation rate of CaCO$_3$, bacterial density has no effect on the growth rate of individual CaCO$_3$ crystals. During the first 1.5 hours after the injection of cementation solution, the crystals grew steadily at the same growth rate even though the bacterial density varied (Figure 6b). The crystals grew to be about 380 ± 40 µm$^3$ by 1.5 hours for all the three bacterial density cases (Figure 6b).

However, the time required for the crystals to finish growing and the final size varied in the three cases. In the low bacterial density case (0.6×10$^8$ cells/ml case), crystal growth continued over 15 hours, which was the longest among the three cases. The average size of the crystals at the completion of crystal growth was about 8000 µm$^3$, which was the largest among the three cases. For the high bacterial density case (2.0×10$^8$ cells/ml case), the overall precipitation rate indicated that the precipitation process completed by around 3 hours (Figure 5b), while individual crystal precipitation rates shows that the process completed by around 10 hours (Figure 6c). The growth of the crystals between 3 and 10 hours is largely due to the dilution of unstable crystals (circled in Figure 6a), which contributed to the growth of larger crystals. This result is consistent with the observation obtained in Wang *et al.* (2019b). The final average size of the crystals was 1800 µm$^3$ after 10 hours. For the very high bacterial density case (5.2×10$^8$ cells/ml), crystals stopped growing by 1.5 hours and the final average size of the crystals was about 400 µm$^3$ after 1.5 hours (Figure 6a and 6b).

**Crystal quantity**

The quantity of crystals formed when the bacterial density was very high (5.2×10$^8$ cells/ml) is compared to when the bacterial density was high (2.0×10$^8$ cells/ml) in Figure 5a and 6a. When the bacterial density was low (0.6×10$^8$ cells/ml), the number of crystals was the lowest among the three cases (Figure 5a and 6a). To illustrate this, the number of crystals in the three cases at each instance when imaging was taken within the first 24 hours were quantified (Figure 7). The CaCO$_3$ crystal concentration represents the quantity of CaCO$_3$ crystals formed per unit volume (i.e. 1 ml). A higher bacterial density resulted in a larger quantity of crystals. When bacterial density was 0.6×10$^8$, 2×10$^8$ and 5.2×10$^8$ cells/ml, the concentration of crystals formed was about 1.5×10$^6$, 7×10$^6$, and 2.1×10$^7$ per ml, respectively, at 24 hours after the cementation solution injection. Bacterial density positively correlated with the number of crystals and the overall crystal growth rate.

The change in the number of crystals with time differed among the three cases. When the bacterial density was 0.5×10$^8$ cells/ml, the number of crystals increased to around 2×10$^6$ per ml soon after the cementation solution injection. When the bacterial density was 2.0 × 10$^8$ cells/ml, the number of crystals increased to around 14×10$^6$ per ml by 3 hours and then decreased to around 7×10$^6$ per ml by around 10 hours. The decrease in crystal number was due to the dissolution of unstable crystal forms. When the bacterial density was 5.2×10$^8$ cells/ml, the number of crystals increased to around 20×10$^6$ per ml by around 3 hours after the cementation solution injection. Dissolution of crystals occurred but was not as obvious compared to when the case when the bacterial density was 2.0×10$^8$ cells/ml.

**Crystal type and dissolution**

As shown in Figures 6 and 7, crystal dissolution occurred in the cases when the bacterial density was either 2×10$^8$ cells/ml or 5.2×10$^8$ cells/ml, but not when the bacterial density was 0.6×10$^8$ cells/ml. To investigate the effects of bacterial density on the dissolution of the crystals, microscope images captured at different instances each of the 12 cementation solution injections are shown in Figure 8. Figures 8a and 8b show the images taken between 0 and 24 hours after the first and the second injections of cementation solution, respectively. Figure 8c shows the images taken at 24 hours after the 3$^{rd}$, 6$^{th}$, 9$^{th}$ and 12$^{th}$ injections of cementation solution.

When the bacterial density was low (0.5×10$^8$ cells/ml in this study), the crystals that were present 24 hours after the cementation solution injection were mainly prismatic, suggesting that the crystals are calcite (Al Qabany *et al.*, 2012; Zhao *et al.*, 2014). These crystals continued growing during the intervals between injections of cementation solution. When the bacterial density was high (2×10$^8$ cells/ml), the crystals which formed after injections of cementation solution were mainly spherical or prismatic. The spherical crystals were relatively unstable and often dissolved, while the prismatic crystals stayed stable. Both the shape and relative solubility of these crystals are consistent with those of calcite and vaterite, respectively (Wang et al., 2019b). When the bacterial density was very high (5.2×10$^8$ cells/ml), bacterial aggregates were observed after the first cementation solution injection and irregular-shaped crystals formed on top of them (Figure 8c). Spherical crystals were also observed. Although this form of CaCO$_3$ precipitates remained present after several initial injections of cementation solution (Figures 8c), they were eventually replaced by more stable forms of CaCO$_3$ crystals. This suggests that when bacterial density is high, the formation of CaCO$_3$ follows the ACC-vaterite-calcite sequence as described in Wang *et al.* (2019b).

A supersaturated state is required for CaCO$_3$ precipitation to occur, meaning that the solution has to contain more Ca$^{2+}$ and CO$_3^{2-}$ ions than could normally be dissolved by the solvent. The supersaturation ratio *S* has been used to quantify the level at which supersaturation induces CaCO$_3$ precipitation, which is defined as:

$$S = \frac{[Ca^{2+}] \times [CO_3^{2-}]}{K_{sp}}$$  Eq. 5

where [Ca$^{2+}$] and [CO$_3^{2-}$] are the concentrations of calcium and carbonate ions, and $K_{sp}$ is the equilibrium CaCO$_3$ solubility product for each experimental temperature (Stumm & Morgan, 1996). A supersaturation ratio that is higher than 1 is required for precipitation to occur.

CaCO$_3$ precipitates can exist as several polymorphs, each of which have different $K_{sp}$ values at different temperatures. At 25 °C, the $K_{sp}$ values of the four main polymorphs of CaCO$_3$ (calcite, aragonite, vaterite, and amorphous $CaCO_3$) are $10^{-8.48}$ $M^2$, $10^{-8.34}$ $M^2$, $10^{-7.91}$ $M^2$ and $10^{-6.40}$ $M^2$, respectively (Plummer & Busenberg, 1982; Brečevic &Nielsen, 1990). Therefore, when $[Ca^{2+}]\times[CO_3^{2-}]$ is below $10^{-8.48}$ $M^2$, no precipitation occurs in any forms; when $[Ca^{2+}]\times[CO_3^{2-}]$ is between $10^{-8.48}$ $M^2$ and $10^{-8.34}$ $M^2$, only calcite precipitates; when $[Ca^{2+}]\times[CO_3^{2-}]$ is higher than $10^{-6.40}$ $M^2$, all forms of $CaCO_3$ can precipitate. On the other hand, after the generation of the different forms of $CaCO_3$, when $[Ca^{2+}]\times[CO_3^{2-}]$ drops to below $10^{-6.40}$ $M^2$ but higher than $10^{-7.91}$ $M^2$, amorphous (ACC) dissolves whereas the other forms of $CaCO_3$ can remain. When $[Ca^{2+}]\times[CO_3^{2-}]$ drops to between $10^{-8.48}$ $M^2$ and $10^{-8.34}$ $M^2$, all forms of $CaCO_3$ dissolve while only calcite remains. When the relatively less stable forms of $CaCO_3$ dissolve, the free $Ca^{2+}$ and $CO_3^{2-}$ can precipitate into the other more stable forms of $CaCO_3$ as long as the supersaturation states are reached. In addition, when multiple forms of $CaCO_3$ can precipitate at the same time, the less stable forms of $CaCO_3$ precipitate more quickly than the more stable forms of $CaCO_3$. Therefore, the ACC-vaterite-aragonite-calcite transformation may occur.

During the MICP process, $Ca^{2+}$ ions are normally present at concentrations in the range of 0.1 M-1.5 M from the beginning of the $CaCO_3$ precipitation process, whereas the initial concentration of $CO_3^{2-}$ is zero (Whiffin *et al.*, 2007; van Paassen *et al.*, 2010; Al Qabany & Soga, 2013; Cheng *et al.*, 2017). A diagram illustrating the relationship between phase transformation and the initial supersaturation state is shown in Figure 9. The concentration of $Ca^{2+}$ was assumed to stay constant at 1.0 M. The supersaturation state is dependent on both the hydrolysis of urea, which increases the concentration of $CO_3^{2-}$, and on the precipitation of $CaCO_3$, which decreases both the concentration of $CO_3^{2-}$ and $Ca^{2+}$. Because bacterial density affects the bulk ureolysis rate, it also affects the supersaturation state, which in turn influences the formation of different phases of $CaCO_3$. When the bacterial density is low, the concentration of $CO_3^{2-}$, which is hydrolysed from urea, increases slowly up to the calcite forming line as shown in Figure 9, after which calcite starts forming. When the concentration of $CO_3^{2-}$ is balanced between the forming lines of aragonite and calcite, only calcite can form. Similarly, depending on bacterial density, the other forms of $CaCO_3$ can also either form or not form, depending on whether the supersaturation state of that type of $CaCO_3$ is researched or not. When multiple forms of $CaCO_3$ precipitate, ACC-vaterite-aragonite-calcite transformation may occur.

**Conclusions**

In this study, microfluidic chip experiments were conducted to investigate the effects of bacterial density on both the kinetics and characteristics of $CaCO_3$ precipitation at the particle scale. Three bacterial densities (0.6 (low), 2.0 (high) and $5.2\times10^8$ (very high) cells/ml counted after the first cementation solution injection) were applied in staged injection MICP procedures. Apart from bacterial density, other experimental parameters including the content and concentration of cementation solution, temperature, and injection flow rate of bacterial suspension and cementation solution were kept constant. Both the overall precipitation rate of $CaCO_3$ and the growth rate of individual $CaCO_3$ crystals were quantified. In addition, crystals characteristics in terms of size, number and dissolution processes were analysed. The main findings of the study are summarised as follows.

When bacterial density is a low ($0.6\times10^8$ cells/ml in this study), the crystals form more slowly than in the higher bacterial density cases, but when sufficient time is given (15 hours in this case), the sizes of $CaCO_3$ crystals were the largest among the three cases. The large crystals could be more efficient in bonding sand with larger particles and larger pore sizes. However, the number of $CaCO_3$ crystals produced is low. Improvement in terms of soil strength may require a certain amount of soil particles to be bonded by $CaCO_3$ crystals. In addition, the time required for $CaCO_3$ to complete the precipitation is long, which implies a long MICP treatment process.

When bacterial density is high ($2.0\times10^8$ cells/ml in this study), the size of $CaCO_3$ crystals formed might be small and unstable, but over time may transform into larger crystals and more stable forms. The crystals formed at the particle contacts should be big enough to efficiently bond the soil particles, which in turn contributes to the strength and stiffness of MICP treated soils. Therefore, when such a high bacterial density is used, the engineering performance efficiency of MICP-treated soils might be better when a longer treatment duration is applied so that the smaller crystals can reprecipitate into large ones.

When bacterial destiny is very large ($5.2×10^8$ cells/ml in this study), the rate of $CaCO_3$ precipitation is increased, but this may be due to the formation of large amounts of the unstable form of $CaCO_3$, ACC. Time-dependent transformation of unstable forms of $CaCO_3$ into more stable forms of $CaCO_3$ occurs. ACC has a lower density compared to $CaCO_3$ crystals, and may transport with flow, or be trapped in soil pores. For soils that have small pores, these unstable crystals may locally clog the soil flow paths and affect the homogeneity of MICP treatment.

Based on solubility, calcite is the most stable form of $CaCO_3$ both physically and chemically, and it has been suggested that the precipitation of calcite is preferred for a permanent stable cementation (van Paassen, 2009). Therefore, when designing an MICP treatment protocol, the effects of bacterial density on the phase transformation and the time for crystals to become stable need to be considered. This study suggests that low bacterial density contributes to the production of stable $CaCO_3$ from the beginning, but the precipitation takes longer to complete. High bacterial density leads to precipitation of less stable forms of $CaCO_3$ first, even though the precipitation occurs faster. A longer treatment time is required for $CaCO_3$ to transform from less stable forms to more stable forms.

The use of microfluidic experiments is useful to assess the ideal bacterial density for various conditions in the field. MICP treatment parameters such as initial bacterial density, bacterial settling time, and injection flow rate of cementation solution all affect the delivered bacterial cell concentration in the soil matrix. Bacterial density directly affects the size and number of $CaCO_3$ crystals formed, which affect the treatment efficiency of MICP for strengthening soils. Correlations between bacterial density and the properties of $CaCO_3$ crystals in terms of number and size from this study could be helpful for the design of MICP treatment protocols for soils with different particle sizes.

**Acknowledgements**

Y.W. would like to acknowledge Cambridge Commonwealth, European and International Trust, and China Scholarship Council, which collectively funded this project. J.T.D. acknowledges the support of the Engineering Research Center Program of the National Science Foundation under NSF Cooperative Agreement No. EEC-1449501. Any opinions, findings and conclusions or recommendations expressed in this manuscript are those of the authors and do not necessarily reflect the views of the National Science Foundation. The authors would also like to thank Dr Fedir Kiskin for proofreading this manuscript.

**Table 1** Summary of bacterial, chemical and injection parameters associated with the microfluidic chip experiments

| Condition No. | Dilution ratio | Bacterial $OD_{600}$ | Injection number | Injection interval (day) |
|---|---|---|---|---|
| 1 | 1:14 | 0.2 | 12 | 1 |
| 2 | 1:5 | 0.5 | 1 | - |
| 3 | 1:2 | 1.0 | 12 | 1 |
| 4 | 2:1 | 2.0 | 1 | - |
| 5 | 1:0 | 3.0 | 12 | 1 |

**Table 2** Summary of the changes in bacterial density during MICP treatment

| $OD_{600}$ Before injection | After BS injection | | After settling | | After CS injection | |
|---|---|---|---|---|---|---|
| | Average ($\times 10^8$ cells per ml) | Derived ($\times 10^8$ cells per ml) | Average ($\times 10^8$ cells per ml) | Derived ($\times 10^8$ cells per ml) | Average ($\times 10^8$ cells per ml) | Derived ($\times 10^8$ cells per ml) |
| 0.2 | 0.95867 | 0.03186 | 1.92 | 0.13 | 0.56933 | 0.09581 |
| 1.0 | 3.92333 | 0.28825 | 6.09 | 0.335 | 2.01333 | 0.16131 |
| 3.0 | 11.9 | 0.61644 | 15.4 | 0.875 | 5.212 | 0.78289 |

Note: BS-bacterial suspension; CS-cementation solution

**Table 3** Summary of the changes in bacterial density during MICP treatment and associated overall precipitation times

| $OD_{600}$ Before injection | Bacterial density ($\times 10^8$ cells per ml) | | | Overall Precipitation | | Reference |
|---|---|---|---|---|---|---|
| | After BS injection | After settling | After CS injection | Time (h) | Rate M/h | |
| 0.2 | 1 | 1.9 | 0.6 | 15 | 0.016 | P1, this study |
| 1.0 | 4 | 6.1 | 2.0 | 3 | 0.083 | P3, this study |
| 3.0 | 12 | 15.4 | 5.2 | 1.5 | 0.160 | P5, this study |
| 1.0 | - | - | - | - | 0.042 | Al Qabany et al., 2012 |

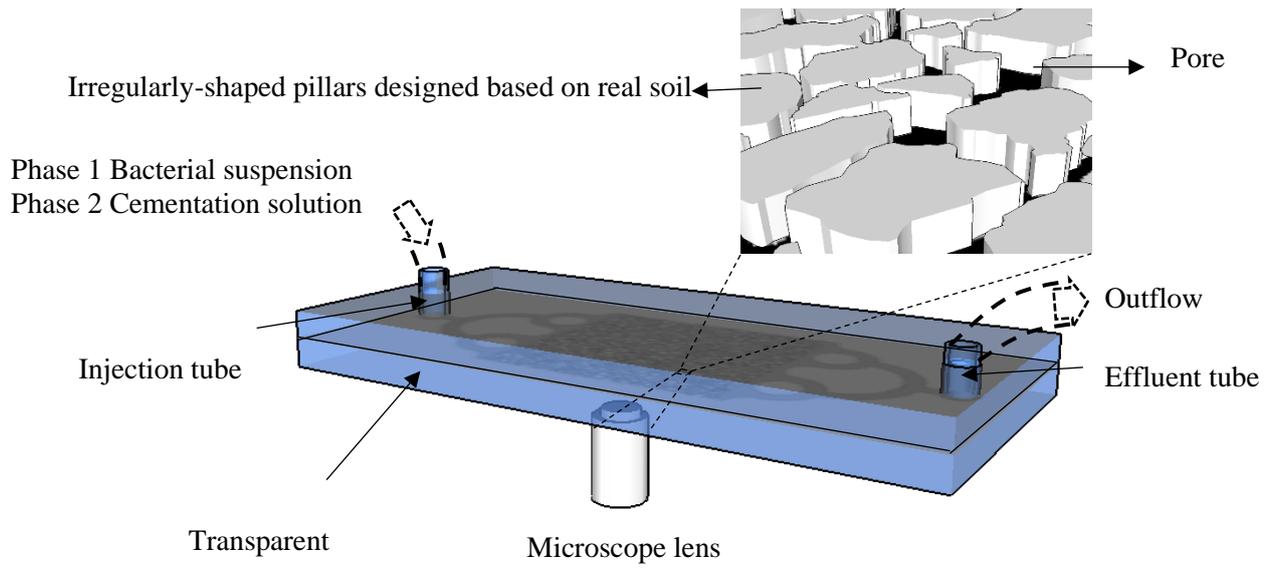

**Figure 1** Schematic of the microfluidic chip experiments (Wang et al., 2019a)

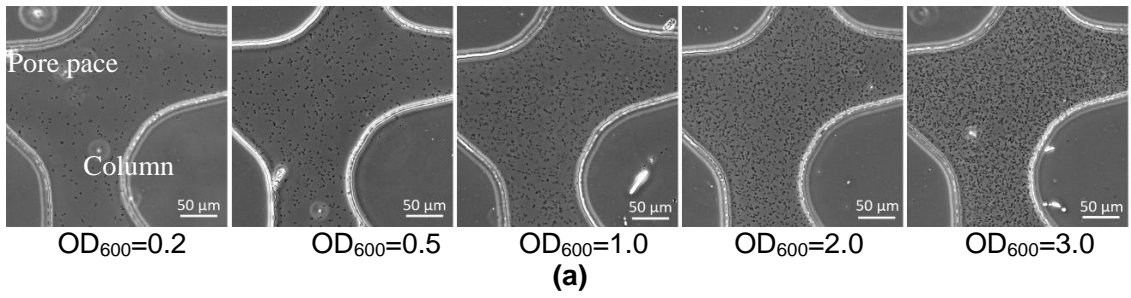

(a)

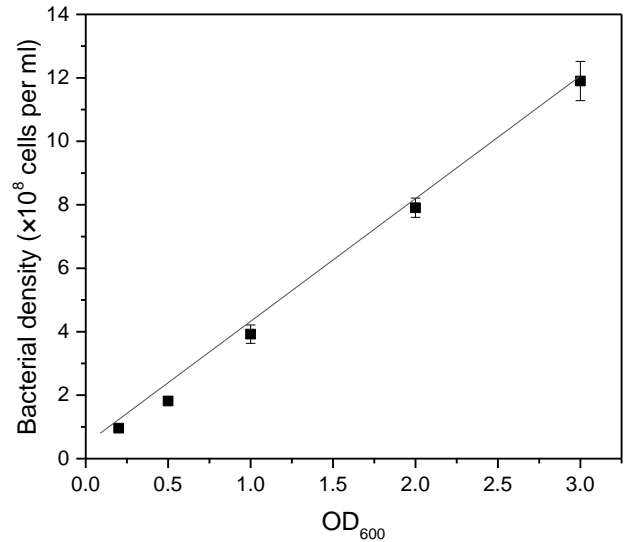

(b)

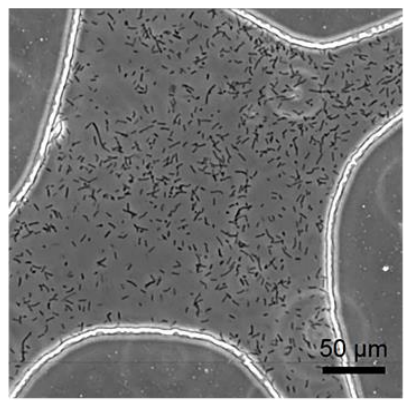 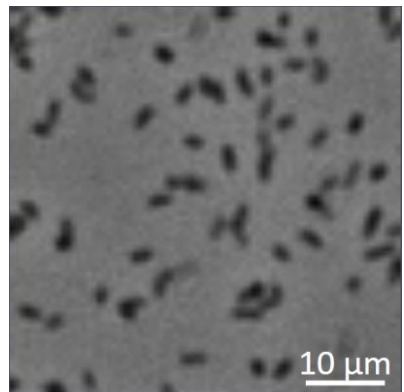

from Wang et al., 2019a       this study

**(c)**         **(d)**

**Figure 2** (a) Microscope images of one pore at the centre of the microfluidic chip taken at ten minutes after the injection of bacterial suspensions with initial $OD_{600}$ values of 0.2, 0.5, 1.0, 2.0 and 3.0; (b) correlations between the initial $OD_{600}$ of the bacterial suspensions and bacterial density at ten minutes after bacterial injection, cell concentration=$OD_{600}\times4\times10^8$, data are presented as mean ± standard error, and each measurement was repeated three times; (c) image of one pore at the centre of the microfluidic chip taken at 2 hours after the injection of a bacterial suspension with initial $OD_{600}$ values of 0.8 showing bacterial size being about 10 μm; (d) magnified image of middle image of (a) showing bacterial size being about 3 μm in this study

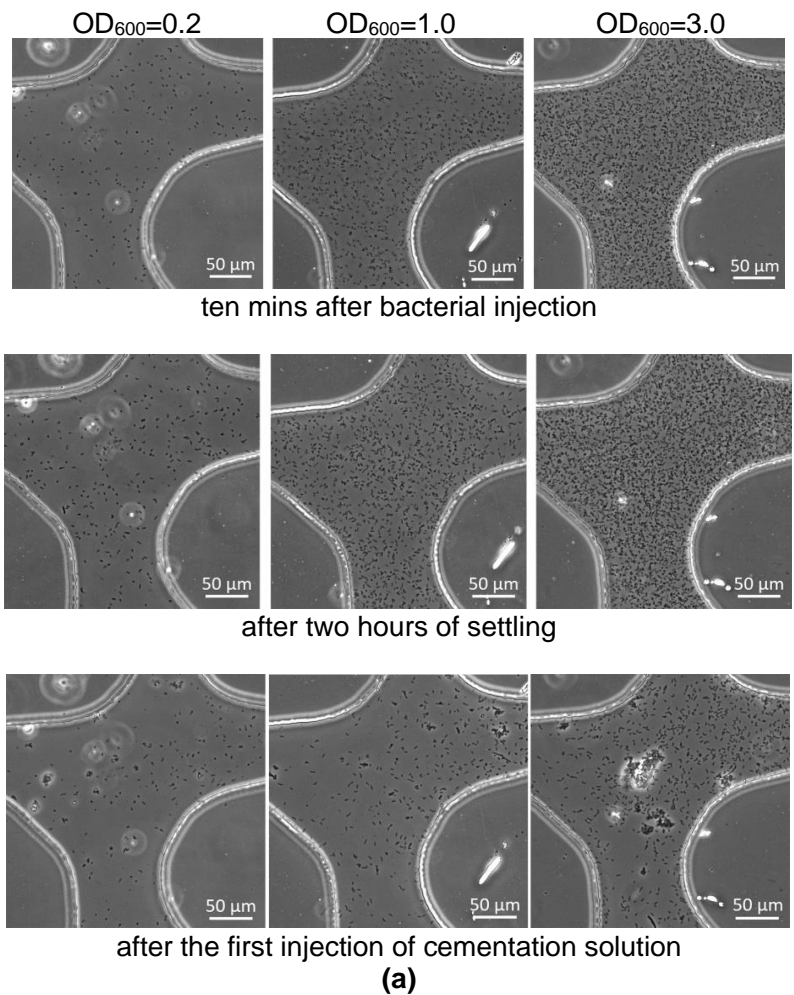

(a)

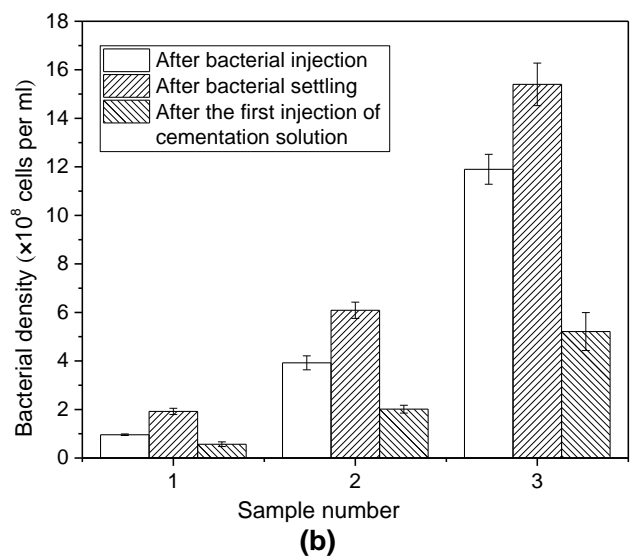

(b)

**Figure 3** (a) Microscope images of one pore at the centre of the microfluidic chip containing bacterial suspesnions with their intial bacteiral $OD_{600}$ were 0.2, 1.0 and 3.0, taken at ten mins after bacterial injection, after two hours of settling and after the first injection of cementation solution; (b) quantification of bacteiral cencentraion in the images

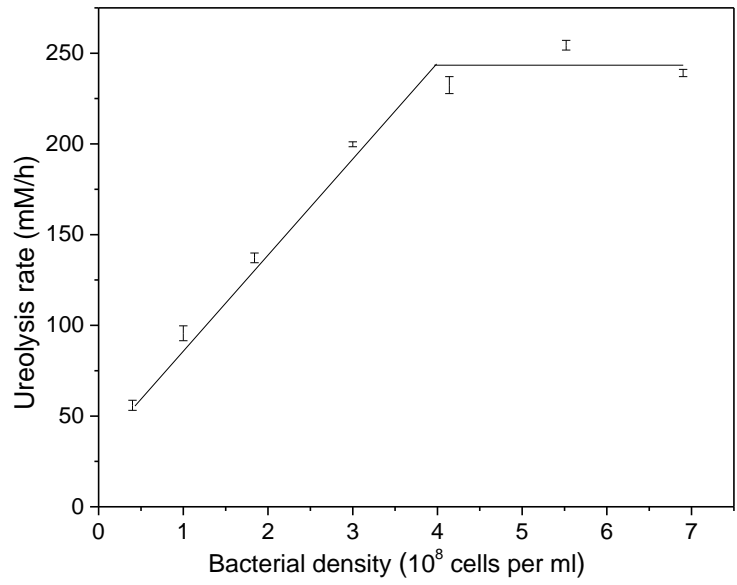

**Figure 4** Ureolysis rate and specific ureolysis rate plotted against bacterial density

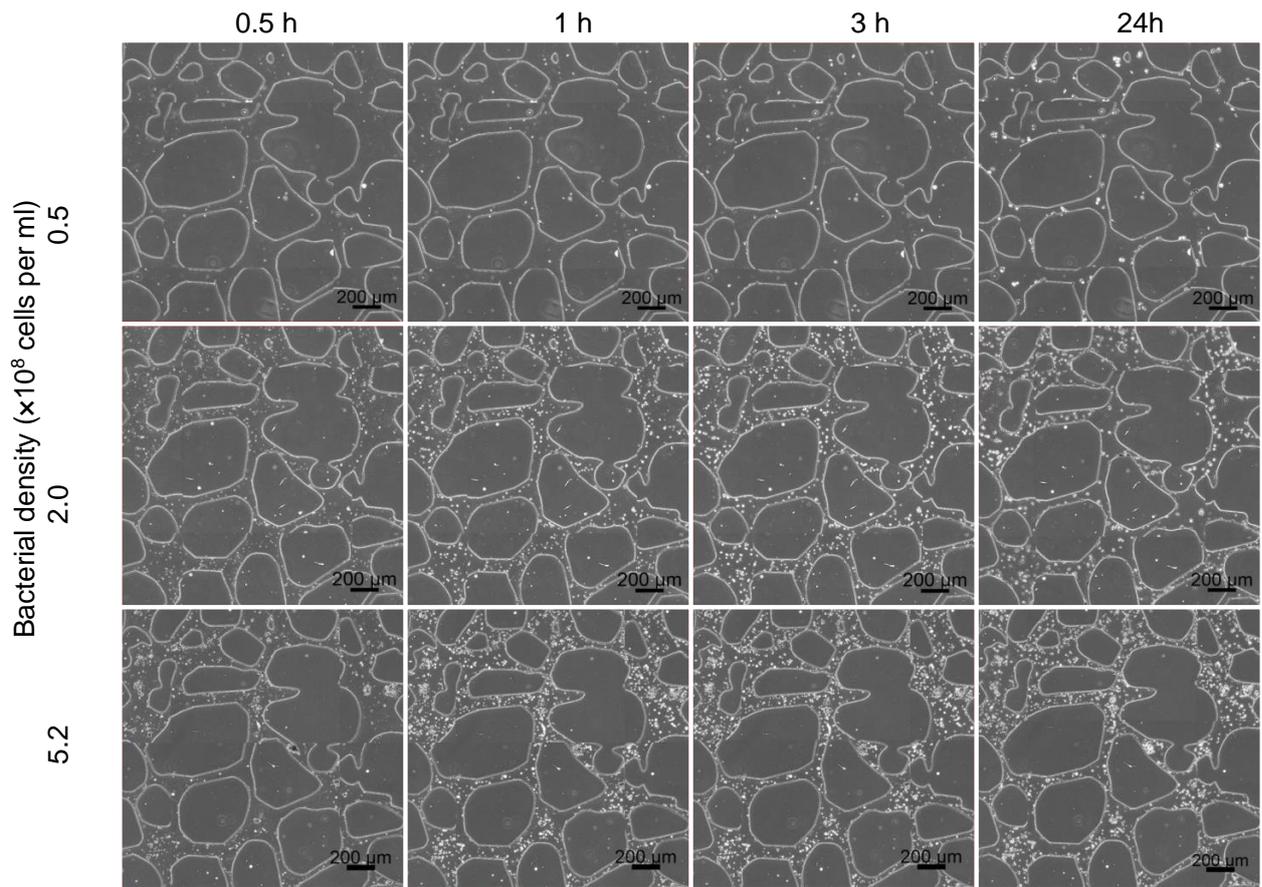

(a)

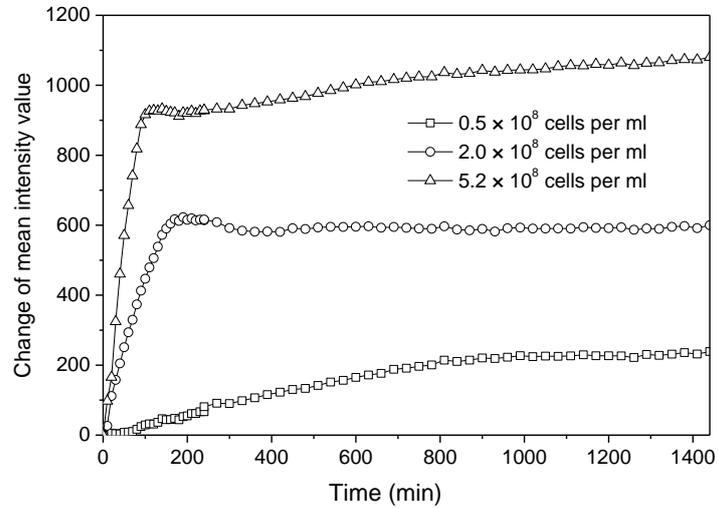

(b)

**Figure 5** (a) Microscope images taken at the centre of the microfluidic chip containing bacterial suspensions at the densities of 0.5, 2.0 and 5.2 ×$10^8$ cell per ml at 0.5, 1, 3 and 24 hours after the first injection of cementation solution; (b) the mean intensity value of the pictures vs. time

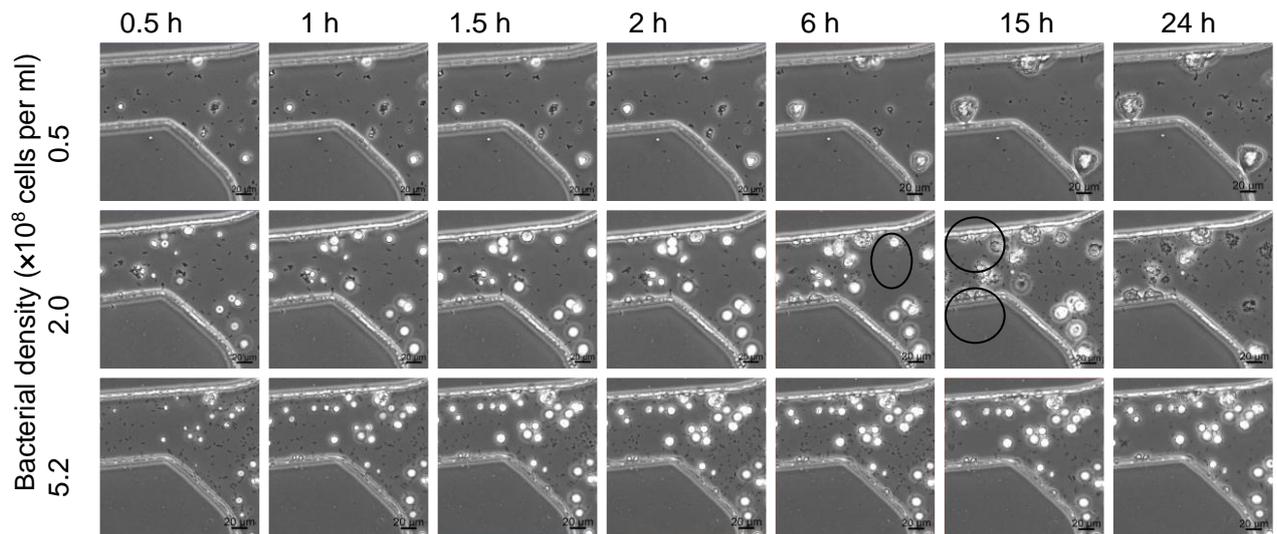

**(a)**

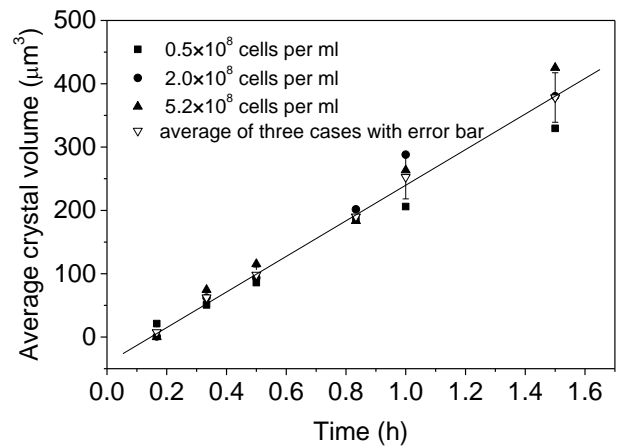

**(b)**

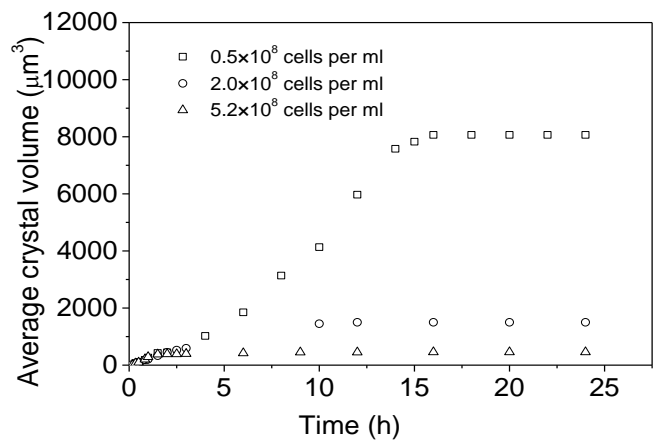

**(c)**

**Figure 6** (a) Microscope images taken at the centre of the microfluidic chip containing bacterial suspensions at the densities of 0.5, 2.0 and 5.2 ×$10^8$ cell per ml taken at 0.5, 1, 1.5, 2.0, 6, 15 and 24 hours after the first injection of cementation solution; (b) the average crystal volume vs. time in the 1.5 hours plotted and their liner fit; (c) the average crystal volume vs. time in the 24 hours after the first injection of cementation solution

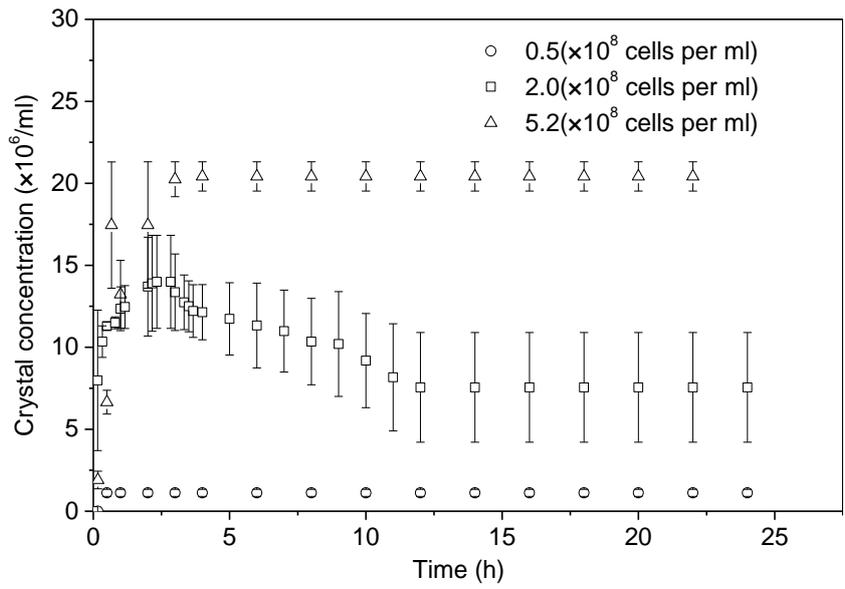

**Figure 7** Scatter plot showing the change in the concentration of $CaCO_3$ crystals with time

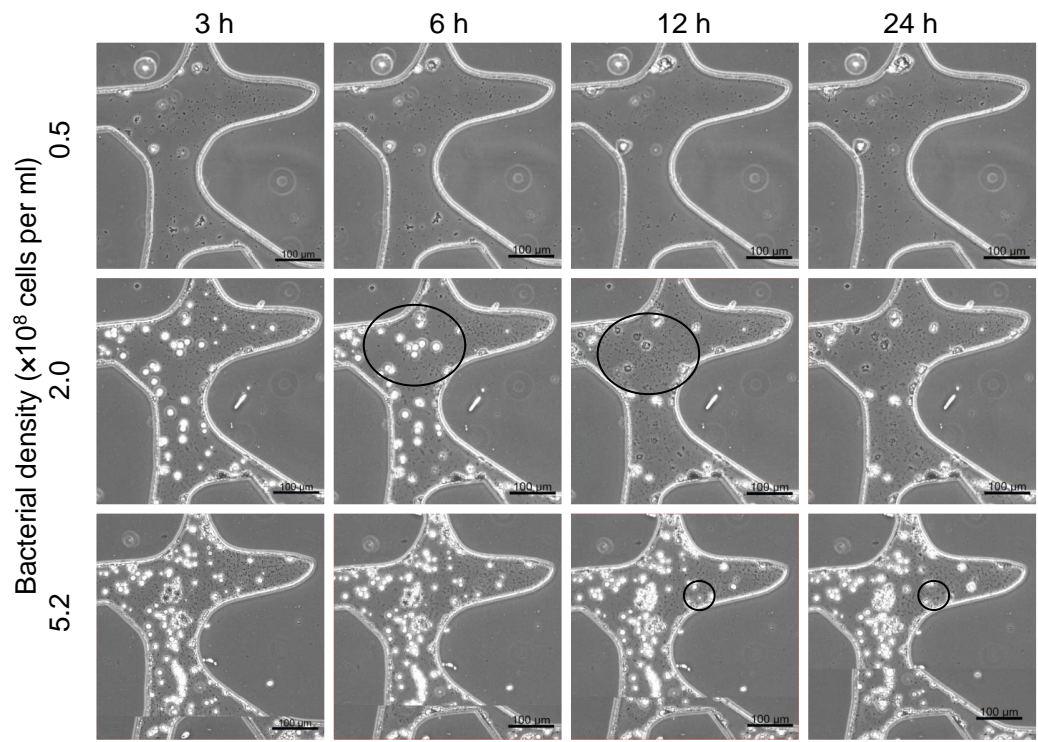

**(a)**

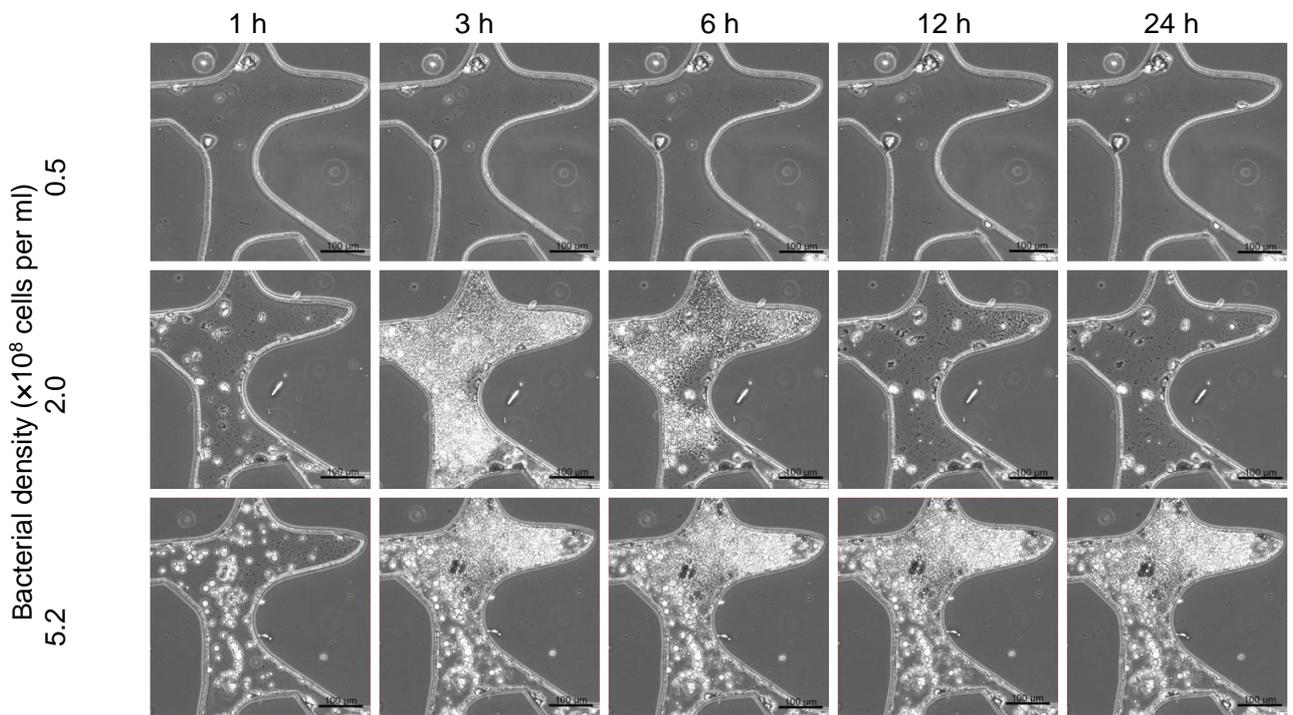

**(b)**

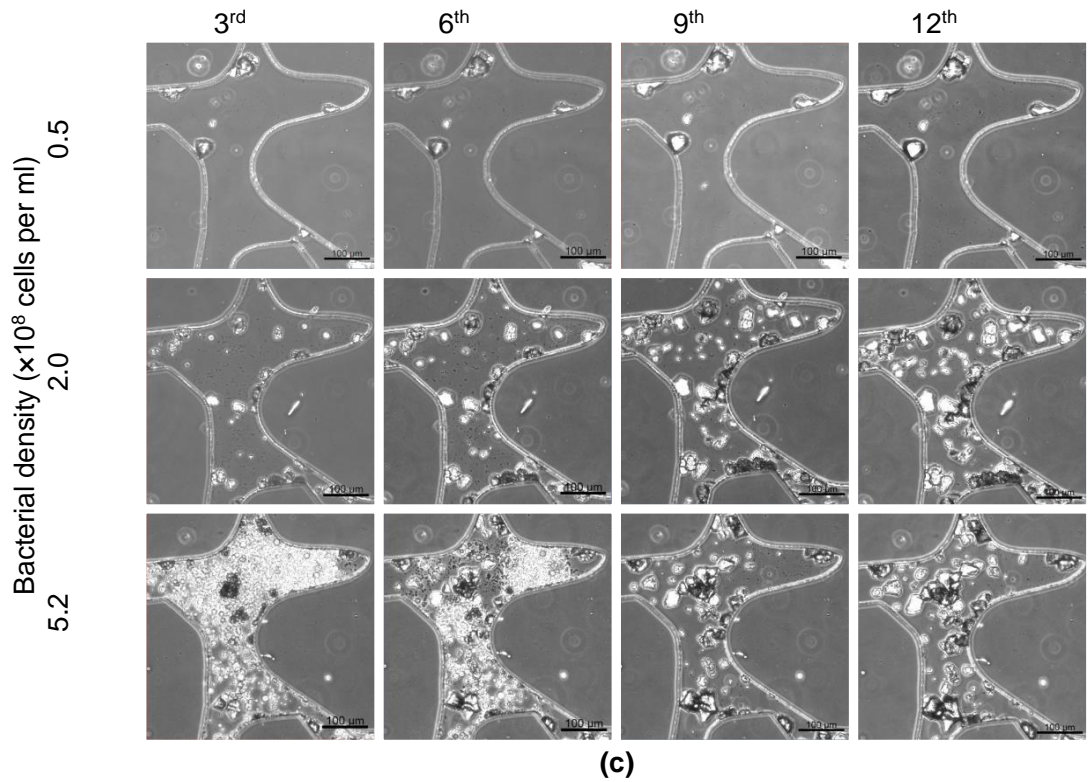

**Figure 8** Microscope images captured at (a) 3h, 6h, 12h, 24h after the first injection of cementation solution, (b) 0h, 3h, 6h, 12h, 24h after the second injection of cementation solution, and (c) 24 hours after the completion of the 3rd, 6th, 9th and 12th injection of cementation solution

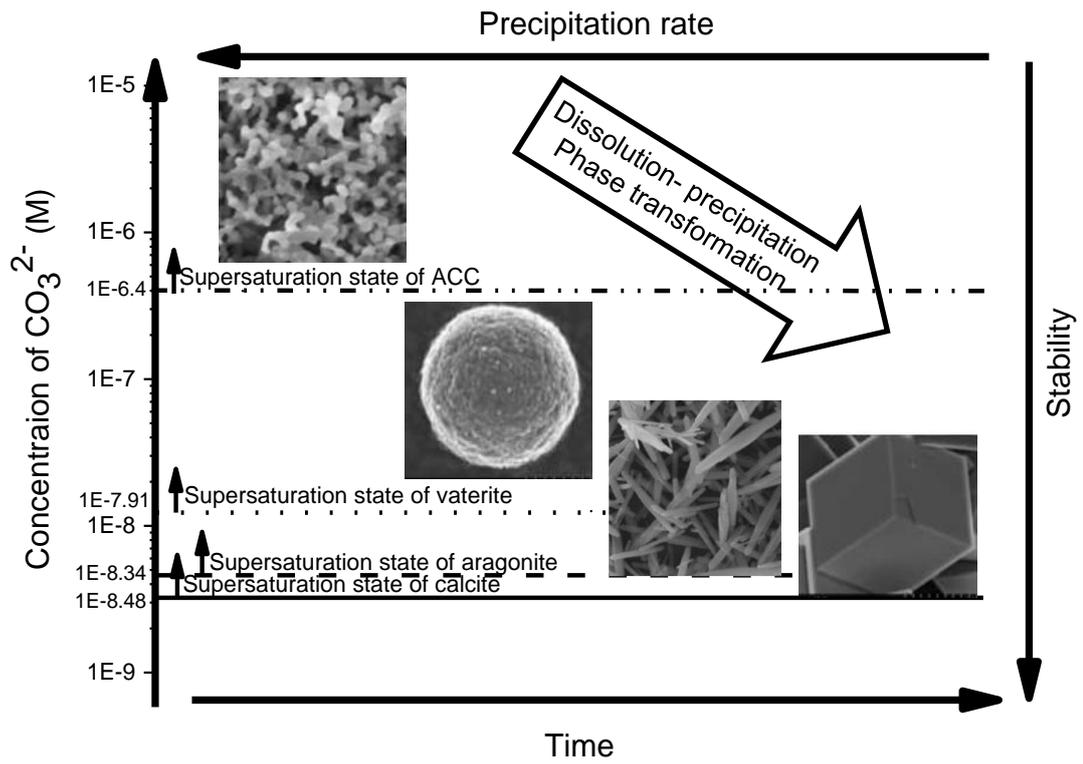

**Figure** 9 Scheme illustrating the precipitation-dissolution and phase transformation, assuming the concentration of Ca2+ is constantly 1.0 M